\documentclass[twoside,fleqn]{article}
\usepackage{espcrc2}
\usepackage[dvips]{graphics}

\newcommand{\bc}{\begin{center}}
\newcommand{\ec}{\end{center}}
\newcommand{\be}{\begin{equation}}
\newcommand{\ee}{\end{equation}}
\newcommand{\bea}{\begin{eqnarray}}
\newcommand{\eea}{\end{eqnarray}}
\newcommand{\del}{\triangle}

\title{Pushing NRQCD to the limit}

\author{Chris Stewart and Roman Koniuk
\address{Department of Physics and Astronomy, 
        York University, \\ 
        4700 Keele St., Toronto, Ontario, Canada M3J 1P3}}
       
\begin{document}

\begin{abstract}
Lattice NRQCD has proven successful in describing the physics of the 
upsilon system and $B$ mesons, though some concerns arise when it is 
used in simulations of charm quarks. It is certainly possible that the 
NRQCD expansion is not converging fast enough at this scale. We 
present some preliminary results on the low-mass breakdown of NRQCD, 
in particular the behaviour of $Q\bar{Q}$ and $Q\bar{q}$ spectra as the 
bare $Q$ mass is decreased well below $1$, with the aim of understanding 
more about the manifestation of this breakdown.  
\end{abstract}

\maketitle

\section{Introduction}
Lattice NRQCD studies of heavy-quark systems have been, on the whole, 
very successful. The predicted spectra for the Upsilon, $J/\Psi$, $B$ 
and $D$ systems, amongst others, agree with the overall structure of 
the experimental spectra, and for the heavier systems, the agreement 
at finer scales, such as hyperfine splittings, is at the percent 
level \cite{ups}.

Trottier's work on the charmonium spectrum \cite{Trottier} indicates 
that the ${\cal O}(v^{6})$-improved NRQCD action 
\emph{decreases} the hyperfine splitting significantly over the 
${\cal O}(v^{4})$ result, rather than taking it towards the 
experimental value. Our own results using similar parameters to 
Trottier confirm this result. This is an indication that NRQCD is not 
converging well for the charm quark.

Yet it is interesting to note that many successful quark model 
predictions of the light-quark spectrum used a \emph{non-relativistic} 
approximation for the light-quark dynamics \cite{quarkmodels}. Here we 
have a highly relativistic system reasonably well described with a 
non-relativistic theory. 

Recently, Liu et al. \cite{WolVal} published work on a model QCD 
theory known as Valence QCD. In VQCD, all z-graphs are removed, and 
the authors were able to make some links between NR quark models and 
the role of z-graphs in standard QCD. We would like to examine the 
behaviour of the low-mass limit of NRQCD, to examine the nature of the
inevitable breakdown of the NR expansion, and see if this can also be 
linked to the remarkable success of NR quark models. In this report 
we present results for the $Q\bar{Q}$ and $Q\bar{q}$ spectra, as a 
function of decreasing heavy quark mass.

\section{Simulation details}
We used an ${\cal O}(v^{6})$-improved NRQCD Hamiltonian, 
\bea
H_{0} &=& \frac{-\del ^{(2)}}{2M_{0}} , \\
\delta H &=& -\frac{c_{1}}{8M_{0}^{3}} \left (\del^{(2)} \right )^{2} 
+ \frac{ic_{2}}{8M_{0}^{2}} \left ({\bf \tilde{\del} \cdot 
\tilde{E} - \tilde{E} \cdot \tilde{\del}}  \right ) \nonumber \\
& & - \frac{c_{3}}{8M_{0}^{2}} \, {\bf \sigma \cdot \left (\tilde{\del} \times 
\tilde{E} - \tilde{E} \times \tilde{\del}  \right ) }
 \nonumber \\
& & - \frac{c_{4}}{2M_{0}} {\bf \sigma \cdot \tilde{B}}+ 
\frac{c_{5}}{24M_{0}} \del^{(4)}  \nonumber \\
& & - \frac{c_{6}}{16nM_{0}^{2}} 
\left ( \del^{(2)} \right )^{2} -\frac{c_{7}}{8M_{0}^{3}} \left \{ 
\tilde{\del}^{(2)}, {\bf \sigma \cdot \tilde{B}} \right \} \nonumber \\
& & -\frac{3c_{8}}{64M_{0}^{4}} \left \{ 
\tilde{\del}^{(2)}, {\bf \sigma \cdot \left( \tilde{\del} \times 
\tilde{E} - \tilde{E} \times \tilde{\del}\right)} \right \} \nonumber \\
& & - \frac{c_{9}}{8M_{0}^{3}} {\bf \sigma \cdot \tilde{E} \times 
\tilde{E}} \,,
\eea
where a $\tilde{\del}$ denotes the use of improved derivative 
operators, and $\tilde{E}$, $\tilde{B}$ are components of the 
improved field tensor. 
The heavy-quark propagator was calculated using the evolution quation
\bea
G_{t+1} &=& \left (1-\frac{H_{0}}{2n} \right )^{n} U^{\dag}_{4} 
\left (1-\frac{H_{0}}{2n} \right )^{n} \nonumber \\
& & \times \left (1 - \delta H \right )  
G_{t} \, .
\eea

For $Q\bar{q}$ mesons, we used the standard tadpole-improved SW action 
for the light quarks, with $u_{0} = \langle Plaq \rangle ^{-4}$ and 
$c_{sw} = u_{0}^{-3}$. The light quark $\kappa = 0.143$, roughly 
comparable to the strange quark mass.

Gauge field configurations were created using a standard tadpole- and 
rectangle-improved action \cite{gauge}. We used an ensemble of *** configurations 
of size $8^{3}\times 24$ at $\beta = 6.8$, which gives a tadpole 
factor of $u_{0} = 0.854$.

\section{Quarkonium}
Results for the $^{1}S_{0}$, $^{3}S_{1}$ and $^{1}P_{1}$ states of 
quarkonium for various values of $M_{0}$ are shown 
in Figure 1. The stabelisation parameter $n$ in Equation (1) was given 
values 
ranging from $2$ to $10$ depending on the value of $M_{0}$. Meson 
operators were smeared at source and sink, and their specific forms 
are detailed in \cite{Trottier}.

\begin{figure}[t]
    \bc
    \scalebox{0.4}[0.4]{\includegraphics{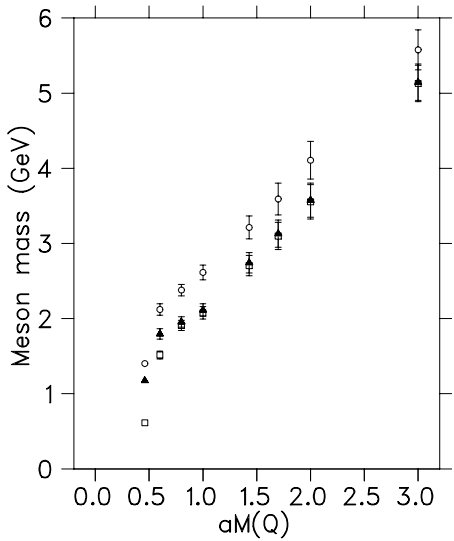}}
    \caption{Quarkonium $^{1}S_{0}$, $^{3}S_{1}$ and $^{1}P_{1}$ states 
    as a function of $M_{0}$.}
    \ec
\end{figure}

The $^{1}S_{0}$ states in Figure 1 are found using the kinetic 
definition of the meson mass,
\be
E_{{\bf p}} = E_{0} + \frac{{\bf p}^{2}}{2aM_{kin}} \,,
\ee
where $E_{{\bf p}}$ and $E_{0}$ are the simulation energies for a 
meson with finite and zero momentum respectively. This method leads 
to the large error bars shown on the data, however the general trends 
are significant here, not the exact energies.

The heaviest $M_{0}$ corresponds to roughly half the bottom quark 
mass, while the charmonium ground state energy of $\sim 3GeV$ 
corresponds to $aM_{0} \sim 1.5$ to $2$. The lightest $M_{0}$ gives a kinetic 
mass close to the energy of an $s\bar{s}$ meson, and so we can assume 
the quarks are well within the relativistic regime at this stage.

Note that the $S$ and $P$ states decrease quite linearly with 
$M_{0}$, until the bare mass drops significantly below $1$, when they 
begin to drop quite suddenly. The hyperfine splitting increases 
significantly at low $M_{0}$, as might be expected from light-meson 
spectroscopy, but the $S-P$ splitting decreases, against observation.
These splittings, with the $^{3}S_{1}$ state taken as the energy 
zero, are shown in Figure 2.

\begin{figure}[t]
    \scalebox{0.4}[0.4]{\includegraphics{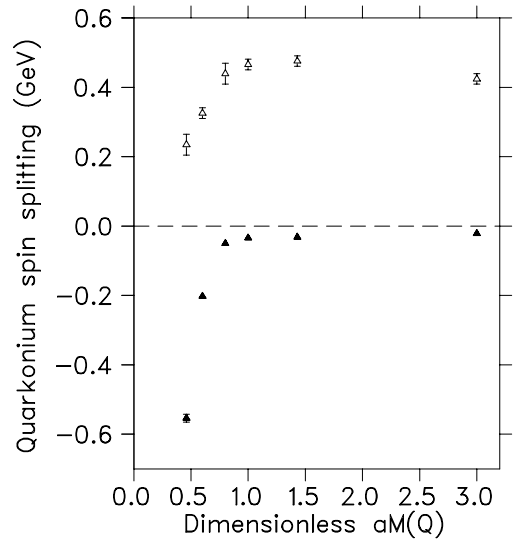}}
    \caption{Quarkonium hyperfine and $P-S$ splittings
    as a function of $M_{0}$.}
\end{figure}

These results for quarkonium indicate the NRQCD action is indeed 
beginning to have difficulties at these low bare masses. The 
situation is worse for the heavy-light spectrum. Figure 3 shows the 
hyperfine splittings for the $Q\bar{q}$ system, and note that for the 
lightest value of $M_{0}$, the splitting is \emph{negative}---the 
$^{3}S_{1}$ and $^{1}S_{0}$ energies are in the wrong order. This is 
obviously a serious pathology.

\begin{figure}[t]
    \scalebox{0.4}[0.4]{\includegraphics{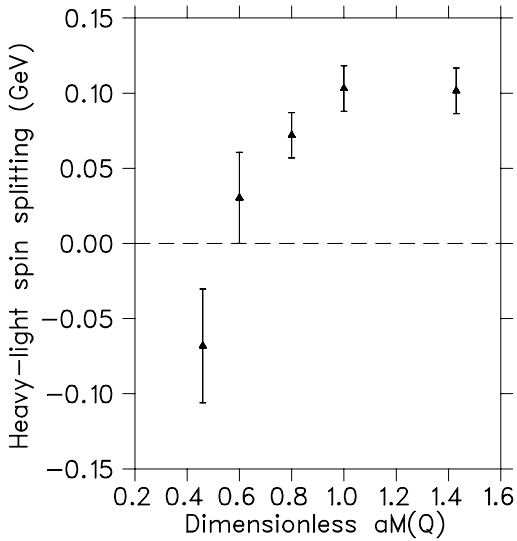}}
    \caption{Hyperfine splitting of $Q\bar{q}$ system
    as a function of $M_{0}$.}
\end{figure}

\section{Discussion}
The results presented here are perhaps not surprising, given that we 
expect the $M_{0}^{-1}$-dependence of the coefficients in the NRQCD 
action will ultimately lead to a breakdown as $M_{0}$ decreases. 
However, we are interested in the exact nature of this breakdown, not 
to see if NRQCD is a viable alternative for simulating light quarks, 
but to illuminate any connections between NRQCD and the successes of 
non-relativistic quark models.

Lewis and Woloshyn \cite{LewWol} have shown that unphysically large
contributions from certain terms in the NRQCD action could be removed by 
subtracting their expectation values from the Hamiltonian. Their 
analysis consisted of a thorough systematic examination of the contributions of 
each correction term in the Hamiltonian. We suspect that the same style 
of analysis, applied to the low-mass breakdown of NRQCD, would flush 
out the terms that suffer the worst pathologies, and may even suggest ways 
they may be strengthened.   

One important issue to be aware of, however, is that the coefficients 
in Equations (1) and (2) are usually given their tree-level value of 
$1$ in NRQCD simulations.
For heavy quarks, tadpole improvement of the gauge field links is 
usually sufficient to account for the most serious corrections 
beyond tree level. The coefficients $c_{i}$ in Equation (2) will, 
in general, have non-trivial $\alpha_{s}(M_{0})$ dependence, 
and this will become more important as the quark mass is decreased.

We would like to thank H. Trottier, R. Lewis, S. Collins and J. Heim 
for their valuable comments. This work is partially funded by the 
National Science and Engineering Research Council of Canada.

\end{document}